\documentstyle[12pt]{article}

\large
\newcommand{\be}{\begin{eqnarray}}
\newcommand{\ee}{\end{eqnarray}}

\begin{document}

\setlength{\baselineskip}{21pt}
\pagestyle{empty}
\vfill
\eject
\begin{flushright}
SUNY-NTG-94/25
\end{flushright}

\vskip 2.0cm
\centerline{\bf Screening of the topological charge in a correlated
instanton vacuum }
\vskip 2.0 cm
\centerline{E.V. Shuryak and J.J.M.
Verbaarschot}
\vskip .2cm
\centerline{Department of Physics}
\centerline{SUNY, Stony Brook, New York 11794}
\vskip 2cm

\centerline{\bf Abstract}
  Screening of the topological charge due to he fermion-induced interactions
is an important phenomenon, closely related with the
resolution of the strong CP and U(1) problems. We study the mechanism of
such screening  in a 'correlated instanton
vacuum', as opposed to the 'random' one. Both scalar and pseudoscalar
gluonic correlators are analyzed by means of an observable
that minimizes finite size effects.  Screening of the topological charge
is established. This allows us to calculate the
$\eta'$ mass without having to invert the Dirac operator. We suggest that
this method might be used in lattice QCD calculations as well.
Our results for the screening of the topological charge are in agreement with
the chiral Ward identities, and the scalar gluonic correlator satisfies a low
energy theorem first derived by Novikov et al. \cite{Novikov-etal}.
We also propose to evaluate
the topological susceptibility in the Witten-Veneziano formula not
in an infinite box in an world $without$ fermions but in an infinitesimal box
in a world $with$ fermions.
\pagestyle{empty}
\vfill
\noindent
\begin{flushleft}
SUNY-NTG-94/25\\
September 1994
\end{flushleft}
\eject
\pagestyle{plain}
\section{Introduction}

Tunneling between topologically distinct sectors of the gauge field,
which is described semiclassically by instantons
\cite{Polyakov_etal,THOOFT-1976},
is known to be one of the major non-perturbative phenomena in QCD.
Their importance was significantly clarified in the last few years.
In particular, a large set of QCD correlation functions was
studied \cite{I,II,III} in the so called 'random
instanton liquid model', RILM.
The results are in surprisingly good agreement
both with  experiment (see \cite{Shuryak_cor}) and lattice data
\cite{Negele_etal}.  They clearly show that
many hadrons (including
e.g. pions and  nucleons) are actually bound by the instanton-induced
interactions, and their properties such as masses (and
even wave functions \cite{Schafer-Shuryak})
can be well reproduced by the simplest instanton-based model.

Secondly, important new results were obtained from the study of
'cooled' lattice configurations. ('Cooling' is a procedure that
relaxes any given gauge field configuration to the closest 'classical
component' of the QCD vacuum.) Not only were they
found to be of the multi-instanton type
\cite{cooling},  but recent work of Chu et al. \cite{Chu_etal_2}
has essentially reproduced the key parameters  of the
'instanton liquid' picture of the QCD vacuum. In particular, it was found
in \cite{Chu_etal_2} that the
mean instanton separation $R\approx
1.1 fm$ and the  typical instanton radius $\rho\approx 0.35 fm$, about 10\%
off the values  suggested
by one of us a decade ago \cite{Shuryak_82}.
They also demonstrated explicitly how hadronic correlators 'survive
cooling' and gave quite convincing arguments showing that
confinement effects play only a relatively minor role.

 In spite of such progress in phenomenology and
lattice studies, a consistent quantitative theory
of instanton-induced phenomena is still missing.
The model mentioned above, RILM,   assumes
 random instanton positions and orientations. It is clearly
the simplest guess, the first step which  ignores
all effects related to
instanton interactions. It seems to be a good  approximation for multiple
phenomena, but there are also important exceptions. In particular, finite
temperature
phenomena related to chiral symmetry restoration transition
cannot be understood without the introduction of
strong correlations between instantons and anti-instantons \cite{T}.
Furthermore, in the phase above $T_c$ these correlations are so strong, that
they  lead to breaking of the liquid into a set of weakly
interacting $\bar I I$ 'molecules'.

   In this paper we discuss some global manifestations
of the quark-induced interaction between
instantons in the $T=0$ case.
It is known that (in the chiral limit) they should screen
{\it large-scale} fluctuations of the topological charge. As it is
a very spectacular manifestation of dynamical
correlations, it is natural to  start our new series of studies
of  instanton-induced phenomena by reporting these results first.
A detailed  report on properties of the correlated instanton vacuum,
including correlation functions and wave functions in particular channels,
will be published elsewhere \cite{SV_next}.

   Before we proceed, let us remind why fluctuations of
the topological charge are
of interest. It is well known that
strong interactions conserve CP to a high accuracy. Two ways to understand
this 'fine tuning' were suggested.
 First, the $\theta-$parameter  may set itself to
{\it zero} due to the axion mechanism
\cite{PECCEI-QUINN-1977}.  A new suggestions by Schierholtz
\cite{Schierholtz} is that $\theta=0$ may happen to be
a phase transition point. In any case, there are all kind of open questions
related to
 QCD at nonzero values of $\theta$, say  whether it is a confining theory
\cite{Schierholtz}.

 The second possibility is that the value of $\theta$ may be {\it irrelevant}
because the total topological charge of the vacuum
is screened completely.
 This  happens if at least one  quark flavor is  massless,
but this does not seem to be the case in the real world.
Recently, a new proposal
was put forward by Samuel \cite{SAMUEL-1992} who argued that the interaction
between instantons may screen the topological
charge  for {\it non-zero} quark masses, below some critical mass value.

  Another important aspect is the famous
 U(1) problem \cite{Weinberg,THOOFT-1986},
related to properties of the $\eta'$ meson.
Phenomenologically, its detailed properties are very important, for example for
understanding of the so called 'proton spin crisis'. Theoretically, a
recent discussion in connection with the Samuel mechanism can be found e.g. in
\cite{DOWRICK-MCDOUGALL,KIKUCHI-WUDKA-1992} in which
the exchange of an $\eta'$ was interpreted as a mediator of
the Debye-type screening. Generally speaking, the 'screening' phenomenon
puts significant constraints on the parameters of this meson. Finite
temperature effects in relation to the interpretation of the $\eta'$ particle
as an inverse screening were discussed in \cite{Ismail}.

A famous approach to the U(1) problem is based on the
Witten-Veneziano formula \cite{Witten-1979,Veneziano-1979}
relating the parameters of the $\eta'$ to the topological susceptibility
$\chi$.
Although it clearly explains qualitative features of the phenomenon,
we cannot be satisfied by it. It is
very difficult to make such type of relation quantitative because the two sides
of the equation 'live in two different worlds' \cite{DOWRICK-MCDOUGALL}.
The parameters of the $\eta'$
meson are obtained from experiment and refer to the real world,
while $\chi$ is calculated
either in the large $N_c$ limit, or in quenched QCD \cite{Jan-Smit},
or in other non-screened theories.
How one matches the unit of energy on both sides is therefore a matter of
some convention.

  No convincing lattice measurements of the $\eta'$ properties in QCD with
dynamical fermions exist so far\footnote{The best one so far
is \cite{Ukawa}, where the
$\eta'$ mass was calculated in {\it quenched}
lattice gauge theory using a rather indirect  method.}.
In this paper we propose a method that does not require
the inversion of the Dirac operator and might therefore be useful for
lattice simulations.

   The results for
the $\eta'$ correlator  in RILM \cite{I,II,III} are also
unsatisfactory: the corresponding
channel has shown to produce too strong $\bar q q$ repulsion.
Besides that, the random ensemble has obviously no screening,
so one cannot get
any constraints related to it. Therefore, in this paper we study an
ensemble of 'interacting instantons', and test whether and how the
corresponding relations hold.
An alternative approach to gluonic correlation functions, both in random and
correlated instanton vacua, is developed in \cite{glue}.

  We suggest that the best way to clarify the issue
is to study the dependence of
{\it topological susceptibility} $\chi(V)$ defined in the {\it sub-volume} V.
Its dependence on V, as well as on
  the quark masses $m_f$, are the main objective  of our work.
 The screening of the
topological charge implies that, {\it for large $V$}, $\chi(V)$
is proportional to the {\it area} of $V$  instead of
$\chi(V)\sim V$,
(see also \cite{SMILGA} where this phenomenon was discussed for the Schwinger
model).
 However, for small volumes one always has that $\chi(V)\sim V$.
The cross-over point between these two regimes defines a screening length,
 identified as the inverse $\eta'$ mass.

In section 2 we define the instanton liquid model and discuss the parameters
used in our calculations. In section 3 the Debije cloud of a topological
charge is studied and in section 4 the volume dependence of the
topological susceptibility is evaluated. These results are analyzed in
the context of correlators that follow from effective field theory
(see appendix A). In section 5 we discuss the scalar
gluonic correlation function which is related to the fluctuations of the
total number of instantons in a given volume. Concluding remarks are made
in section 6.

\renewcommand{\theequation}{2.\arabic{equation}}
\setcounter{equation}{0}
\vskip 1.5 cm
\section{The model}
\vskip 0.5 cm
The basis of this work is the
'interacting instanton approximation' (IIA), formulated as a particular
statistical model amenable for numerical simulations
\cite{Shuryak_1988,Shuryak_1989,SHURYAK-VERBAARSCHOT-1991} or
  analytical studies \cite{Diakonov_Petrov,NVZ}).
The partition function for $N/2$ instantons
and $N/2$ anti-instantons is generally given by
\be
Z = \int \prod_{I=1}^N d\Omega_I dz_I d\rho_I \mu(\rho_I)
\prod_f^{N_f}\det(T+im_f)
\exp[-\beta(\rho\Lambda) \sum_{I < J} S^{int}_{IJ}],
\ee
where $\Omega_I, z_I$ and $\rho_I$, denote the orientation, position and
the size of pseudoparticle $I$, respectively. The size distribution
$\mu(\rho)$ contains a factor $\sim \rho^{b-5}$ (with
the standard beta-function parameter $b = \frac{11}{3} N_c -
\frac 23 N_f$) due to
Jacobian of the transformation to collective coordinates and the
leading quantum corrections ($\beta(\rho\Lambda) = -b \log(\rho\Lambda$).
The QCD parameter $\Lambda$ will be determined phenomenologically.

The main dynamical effects we are going to study are  due to
 the determinant of the Dirac operator.  In the
subspace of the zero modes it can be expressed in terms of the overlap matrix
elements $T$, which for the 'streamline' gauge fields
\cite{VERBAARSCHOT-1991} are given by
\cite{SHURYAK-VERBAARSCHOT-1992A}
\be
T_{AI}(\rho_I \rho_A)^{\frac 12}
=(\rho_I \rho_A)^{\frac 12}\int d^4x \psi_0^{\dagger A}(x) i\hat D
\psi_0^I(x)=
\frac 12 Tr \left ( \tau_\mu^+ R^{IA}_\mu U_I^{-1} U_A \right )  F(\lambda),
\ee
where the scalar function $F(\lambda)$ is defined by
\be
 F(\lambda) =
6\int_0^\infty {dr r^{3/2} \over (r+1/\lambda)^{3/2} (r+\lambda)^{5/2}}.
\ee
It depends on the conformally invariant combination
$\lambda$
\be
\lambda  = \frac 12\left(\tilde R^2 +\frac{\rho_I}{\rho_A}+
\frac{\rho_A}{\rho_I} \right ) +
\frac 12\left  ((\tilde R^2+\frac{\rho_I}{\rho_A}+\frac{\rho_A}{\rho_I} )^2
-4\right)^{\frac 12},
\ee
where $\tilde R=R/\sqrt{\rho_I \rho_A}$.
Asymptotically, for large $R/\rho$, the $T_{AI} \sim 4\rho^3/R^3$ (the
geometric average of $\rho_I$ and $\rho_A$ id denoted by $\rho$).


   Considering the interaction due to gauge fields, let us first mention that
we deviate from previous works\footnote{They used
  variational trial functions for the gauge fields
known as the 'sum' or
 'ratio' ansatz.}
by using the interaction derived in \cite{VERBAARSCHOT-1991}
from the  so called 'streamline equation' \cite{Yung}. They
correspond to the true bottom of the
valley for
the $\bar I I$-configurations.
 This  interaction differs from
the previously used ones in one important aspect:
for one relative orientation the repulsive 'core' at small
distances is gone.
  Our numerical simulations have
confirmed earlier suspicions that
it leads to an  'overcorrelated  liquid' of close $\bar I I$ pairs.
It is still possible, that one may get a phenomenologically acceptable
ensemble by a significant increase in the density of instantons. However,
to spend a major portion of computer time simulating pairs that
are not even semiclassical fluctuations is not practically
possible.

  Thus, it seems that the original hopes to stabilize the ensemble
at the purely classical level \cite{Diakonov_Petrov} are not
fulfilled. Presumably quantum effects (especially
subtraction of  perturbative contributions, relevant for close
instanton-anti-instanton pairs with a strongly attractive interaction) will
generate the effective repulsion, and
two recent developments should be mentioned in this context. First,
Diakonov and Petrov calculated the instanton interaction from the semiclassical
quark scattering amplitude \cite{Diakonov-Petrov-1993} and found a short range
repulsion. However,
at the moment it is not yet clear to what extent this interaction
is of relevance to the present problem.
Second, some quantum corrections,
discussed in \cite{Langfeld-Reinhardt} using the scale anomaly relation,
also lead to some effective repulsion.

Lacking a detailed understanding of these effects,
we have introduced
a {\it phenomenological repulsive core}\footnote{Let us point out an analogy to
the famous problem of nuclear matter saturation:
although there is no doubt for the existence of a repulsive core for NN forces
and its decisive role in the problem,
its properties and physical nature is debated even today, after decades
of investigations!} to the streamline action $S^{SL}(\lambda)$
\be
S^{int}_{IA} &=& S^{SL}(\lambda) + \frac A{2\lambda^4}
{{\rm Tr} {\bf 1}_2 U_I^{-1}U_A {\bf 1}_2 U_A^{-1}U_I },\\
S^{int}_{II'} &=&  \frac A{2\lambda^4} {{\rm Tr}
{\bf 1}_2 U_I^{-1}U_I' {\bf 1}_2 U_{I'}^{-1}U_I},
\ee
for pseudoparticles  of the opposite and identical kind, respectively.
(In the latter case the  classical interaction is absent and only the
hard core is included.) The strength of the core is denoted by a free parameter
$A$, and the color traces are needed to
to reduce
the magnitude of the interaction if two pseudoparticles belong to
different SU(2) subgroups of $SU(N_c)$.

  In principle, $A$ determines the total instanton density
in terms of $\Lambda^4_{QCD}$. Unfortunately, it is not known with sufficient
accuracy, and therefore it is logical to do the same thing as on the lattice,
namely tune those two
 parameters to some measured masses, and then consider others
as predictions. This strategy will be used in a subsequent work on
hadron spectroscopy in the IIA.

  However, in this paper, related to only a limited number of issues, we adopt
a
simpler normalization prescription. For all masses, number of colors or flavors
we  (i) fix the total
density of the instantons at $n=1 fm^4$, roughly corresponding to
the standard gluon condensate value and to lattice measurements;
  (ii) the repulsive
core strength is fixed to be $A = 128$, for which
 we obtain
a {\it quark condensate} that is close to the empirical
value. (Note that in this region the dependence on $A$ is relatively weak.)

We simulate the partition function with a
Metropolis algorithm for an ensemble of $N/2 =32$
instantons and and an equal number of anti-instantons in a box of size
$(2.38 fm)^3 \times 4.76 fm$. We average over typically 1,000
statistically independent configurations.

\vskip 1.5cm
\renewcommand{\theequation}{3.\arabic{equation}}
\setcounter{equation}{0}
\section{Screening of the topological charge}
\vskip 0.5 cm
 Let us start with the evaluation of
 the topological charge
\be
Q(l_4) = \int_{H(l_4)} d^4x \frac{g^2}{32\pi^2} F \tilde F(x),
\ee
in a subvolume ('slice' of the box)  $H(l_4)\equiv L^3\times l_4$.
It
 is not an easy task on a lattice,
but for our   instanton ensemble
this simplifies to just counting instantons and anti-instantons
with centers in the subvolume.

  Our aim is to study the details of the screening phenomenon such as
the {\it size} and {\it shape} of the
corresponding 'Debye cloud'.  Let us take
the center of one of the anti-instantons
as the origin of the coordinate system and measure the charge
distribution in the rest of the system.
 In Fig. 1 we show results of such studies for $N_c =3$ and $N_f=2$ (both u,d
quarks
with equal masses, given in the label of the figures).
Since the total charge in the
box is set to zero,
the 'compensating charge' contained in the slice $H(l_4)$
for large enough $L_4$ should approach 1
(the anti-instanton in the center is not counted).

  Our data (points) do show a strong screening phenomenon, namely that
most the
 compensating charge is contained in rather small slice of the box, with
 a width of only about $1/3$ fm.
That means that the correlations between instantons and anti-instantons,
responsible for screening, are actually
very short range ones. The relevant
hadronic excitation, the pseudo-scalar flavor singlet channel,
should therefore be relatively heavy.
Thus, our ensemble solves
the $U(1)$ problem at least qualitatively.

 One more qualitative phenomenon to mention here is the
  observed charge oscillations around the central
charge. Those are well seen for $m=0$, and fade away for larger masses.
The oscillations are clearly non-statistical, and their appearance
 suggests that in the 'instanton liquid'
there exist correlated clusters of various
sizes. Let us recall in this connection, that in many ordinary liquids
one finds local order characteristic to a crystalline phase, and that clusters
can be as big as containing hundreds of atoms. This interesting topic
clearly deserves further studies.

  Doing a more quantitative analysis, we have fitted these data by
  the function
\be
Q(l_4) = \alpha^2(1 - \exp(-m l_4)) + (1-\alpha^2)\frac {l_4}L
\label{fit}
\ee
The fraction $\alpha$ is the
magnitude of the 'compensating charge', if $\alpha = 1$ the
charge is  screened completely. The second term corresponds to the
remaining charge which is assumed to be distributed homogeneously over
the box. The first term is obtained by integrating the correlation
function (\ref{eff10}) for fixed $x_4 - y_4 = l_4$. In the effective low energy
field theory (discussed in appendix A) the mass $m$ is identified
as the $\eta'$ mass.

 The obtained values
for $\alpha$ are 1.0, 0.99, 0.96 and 0.80 for the 4 masses
 used ($m=$ 0.0, 0.1, 0.2 and 0.4
in units of $fm^{-1}$), with statistical errors about 0.05.
Thus, the non-screened charge $(1-\alpha^2)\sim m^\delta$ with the power
$\delta=1$ to 2. This conclusion should be compared to
the conventional wisdom, according to which
it should be proportional to the lightest quark mass. We do not know
whether there is a discrepancy, or that it is due to finite
 size effects (the $1/L$ term
in (\ref{fit})). In the scenario proposed by
Samuel, the effect should vanish at a non-zero critical mass
$(1-\alpha^2)\sim (m-m_c)^\delta$. Our results lead to an upper limit
of about $m_c < 20 MeV$.

The value of the second parameter, $m_{\eta'}$ equals
673, 623, 555 and 518 MeV, in these 4 cases, respectively\footnote{
We remind the reader that the 'MeV' makes sense in those un-physical
theories only after we have defined them, rather arbitrarily, by keeping
the density of instantons fixed at some value. Comparison to lattice
simulations is possible, but one has to keep the same convention.
Th physical situation will be discussed below. }.
 The result
is  that singlet mass {\it decreases} with the
current quark mass and suggests that the screening length
becomes larger if one is
increasing {\it all} quark masses.

\renewcommand{\theequation}{4.\arabic{equation}}
\setcounter{equation}{0}
\vskip 1.5 cm
\section{Fluctuations of the topological charge in a subvolume}
\vskip 0.5 cm

   In this section we switch to more conventional observables,
which have been discussed  in both theoretical and lattice literature.
They are closely related to what was done above, but at the same time they
do not demand  the charge to be localized.

Although   the average value of the
topological charge in some volume is, of course, zero, one may
study its {\it fluctuations} $\langle Q^2 \rangle$.
More specifically, we study the
dependence  on the  the finite volume $V_4$, its surface $A_3$, etc.,
\be
\langle Q^2\rangle_V= C_V(m)\,V_4+ C_A(m)\,A_3.
\ee
(The subscripts refer to the dimensions of the manifolds).

   We have already mentioned, that the
attention in
 literature   has been focused on the {\it large}
volume limit of that quantity, the topological susceptibility
\be
\chi = \lim_{V_4\rightarrow\infty} {\langle Q^2\rangle_V \over V_4}.
\ee
The dependence of $\chi$ on
the quark masses $m$ is the same as for unscreened
charge $1-\alpha^2$ discussed above: the standard scenario with the
'screening of the topological charge' implies that
$\chi\sim m$ at small $m$, while in
Samuel's scenario $\chi=0$ for all quark masses
below some critical value  $m < m_c$.

Our idea is to study not only the limiting value of the topological
susceptibility but its dependence on the {\it subvolume}.
The particular behavior to be observed is very informative,
and can be used to extract additional information about vacuum properties
(and hadronic spectroscopy).
Specifically, as we try to separate {\it volume}  from {\it surface}
effects, we take a segment of the
hypertorus of variable length $l_4$ along the 4-axis, i.e. the slice $H(l_4)$
discussed in previous section.
Its total volume $L^3 l_4$ is proportional to $l_4$,
while its surface area $L^3$ is independent of it. Because our box size in the
4-direction is  $L_4=2 L$, the effective range of $l_4$ is $0<l_4<L$.

Furthermore,  as we have seen in appendix A, the topological correlation
function $\Pi_P(x-y)$ (see (\ref{eff10}))
can be connected with parameters of the SU(3)-scalar pseudoscalar mesons,
the famous $\eta'-$particle and the corresponding excited states.

The fluctuations of the charge $Q(l_4)$ in the box $H(l_4)$ are obtained
by integrating the correlation function (\ref{eff10}):
\be
K_P(l_4)\equiv  \langle Q(l_4)^2\rangle
       = L^3\int_{L^3} d^3 x \int_{-l_4}^{l_4} dt(l_4 - |t|)
\Pi_P((\vec x^2 + t^2)^{\frac 12}). \ee
This  integral vanishes for $l_4 \rightarrow \infty$
if {\it any } quark becomes
 massless\footnote{A similar cancellation takes place in many
cases, e.g. in (2-d) Schwinger model (see
 \cite{SMILGA} for details).} (see eq. (A.15) in appendix A).
For finite $l_4$ this integral can be done as well;
for sufficiently large $L\gg 1/m$ the range of the
spatial integration can be extended to infinity
with high accuracy $O(\exp(-L m_{\eta'}))$ and the result is
\be
\langle Q(l_4)^2 \rangle=
L^3n\left [ l_4
+\cos^2\phi \frac {m_t^2}{m_{\eta'}^2}
\left(\frac{1-\exp(-m_{\eta'} l_4)}{m_{\eta'}}-l_4\right ) +
\sin^2\phi \frac {m_t^2}{m_{\eta}^2}
\left(\frac{1-\exp(-m_{\eta} l_4)}{m_{\eta}}-l_4\right)\right ].
\nonumber \\
\label{k_p}
\ee
For $l_4 \rightarrow 0$ we find
\be  \langle Q(l_4)^2 \rangle = n L^3 l_4.
\ee
One can easily convince oneself that this result is independent of the
specific shape of the subvolume, and in general we have
\be
\lim_{V\rightarrow 0}
\frac{\langle Q(V)^2 \rangle}{V}= n.
\ee
Together with eq. (\ref{eff11}), this results in the following
formula for the $\eta'$ mass
\be
m_{\eta'}^2+m_{\eta'}^2-m_{K}^2=
\frac{2N_f}{f^2}\lim_{V\rightarrow 0}\frac{\langle Q(V)^2 \rangle}{V},
\ee
which is our analog of Witten-Veneziano formula,
 with the topological susceptibility
replaced by {\it local} fluctuations of the topological charge. It should hold
for volumes in the window between
 that of a single instanton $O(\rho^4)$ and the inverse of
their density $O(1/n)=O(R^4)$.

In terms of the flavor singlet spectral function,
the correlation function (\ref{k_p}) implies that neither the contribution of
the $\iota$ nor the contribution of the continuum have been taken into account.
We only calculate the non-perturbative contribution to the correlation function
$\Pi^{NP}_P= \Pi_P - \Pi_P^{pert.}$ which implies that the contribution
of the continuum and the resonances appear with opposite signs in
$\Pi^{NP}_P$. In particular we will find a cancellation between the $\iota$ and
the continuum. In the fits to be discussed below we included the contribution
of the  continuum, and in all cases we found that the best fit was obtained
for zero coupling of the continuum. From now on we do not include it in
our discussion.

Our results for
the fluctuations of the topological charge, $K_P(l_4)$, in a slice with
length $l_4$   are shown in Figs. 2, 3 and 4.
The first qualitative point we would like
to make is that the data points clearly show
 the expected transition from
a {\it linear dependence} on $l_4$ (for small box size)  to a {\it
constant} for larger boxes (in case the quark mass is zero).
These data point deviate drastically from the
full line corresponding to randomly positioned
instantons\footnote{ The functional form of this curve,
which follows immediately from
the binomial distribution of the instantons, is given by
$N (l_4/L_4)(1-(l_4/L_4))$.},
the behaviour that holds for the RILM used in \cite{I,II,III}.
Thus, the topological charge is in fact {\it screened} and its fluctuations
become a surface rather then a volume phenomenon, as soon as the
quark-induced interactions are included in the statistical sum.

 In the upper figure of
Fig. 2 we compare the cases with {\it one}
and {\it two} massless quarks, for $N_c = 2$.
For massless quarks the correlation function (\ref{k_p}) reduces to
\be
\langle Q(l_4)^2 \rangle_{\rm chiral}= \frac{1-\exp(-m_{\eta'} l_4)}
{m_{\eta'}}.
\ee
The dashed curves show a fit of this function to the data points. For
$N_f =1 $ and $N_f =2$ we find $m_\eta' = 535 MeV$ and
$m_\eta' = 756 MeV$, respectively. The ratio of the
masses is 1.41 which is right on top
of the theoretical expectation that $m_\eta' \sim \sqrt{N_f}$ (see
(\ref{eff11})).
If we study the susceptibility $\langle Q(l_4)^2 \rangle$ as a function of
the quark mass, which effectively reduces the number of flavors, we
therefore expect
on the one hand that the $\eta'$ mass decreases, whereas, on the
other hand, it increases because of the explicit contribution of the quark
masses.

The quark
 mass dependence of the topological correlator is studied in Fig. 3.
In this case the number of colors is 3 and we consider two flavors with
equal mass (same instanton configurations as in Fig. 1).
For the quark mass we refer to the label
of the figure.

In the case of equal quark masses, the fitting formula (that follows
from eq. \ref{eff10}) is
\be
\langle Q(l_4)^2 \rangle= n L^3\left
(\alpha^2 l_4(1-\frac{l_4}{2L})+ (1-\alpha^2)
\frac{1-\exp(-m_{\eta'} l_4)}{m_{\eta'}}\right ).
\ee
The value of $\alpha$ obtained from a least square fit is
0.0, 0.0, 0.0, 0.19 and $m_{\eta'}$ equals 580, 520, 430 and 340 MeV
for quark masses of 0.0, 20.0, 40.0 and 80.0 MeV, in this order.
As we already observed in previous section, our results set a limit on
the existence of a critical mass below which we have complete screening.
Again with the disclaimer that the individual parameters are not well
determined by the fit. Only the height of the curves is well-determined.
Our values for the $\eta'$ mass are somewhat less than in previous section,
but show the same trend that it decreases for not too large quark masses.

Finally, in Fig. 4,
we show results for realistic quark masses\footnote{Of course, for any
finite volume simulation the value of $m_f$ cannot be taken arbitrary small:
the minimum possible value
of $m$  is the point where the $m-$derivative of
the condensate vanishes. For feasible volumes we work with
$m_f =0.05 \approx 10 MeV $, being small compared to
hadronic masses but
still considerably larger than the  light
quark masses in the real world.},
$m_u = m_d = 10 MeV$
and $m_s = 150 MeV$. The dashed line corresponds to effective theory with
realistic meson masses and $\eta-\eta'$-mixing angle. The full line is a fit
of eq. (\ref{k_p}) to the data. Both the pion and the kaon mass are found to be
zero, whereas $m_\eta' = 605 MeV$. Also in this case
the individual parameters
are not well determined by the fit. In particular, the fitted curve is very
insensitive to the value of the strange quark mass.
In fact, the best fit
misses the strange quark contribution to the $m_{\eta'}$.
Note that in naive chiral perturbation theory
this meson should possess a strangeness-related mass
$m^{strange}_{\eta'}= m_{\eta}/\sqrt{2}= 388 $ MeV. The sum of these two
numbers are right in the correct region.

\vskip 1.5 cm
\renewcommand{\theequation}{5.\arabic{equation}}
\setcounter{equation}{0}
\section{Compressibility of the instanton liquid}
\vskip 0.5 cm
Let us now proceed to studies of a
related gluonic correlation function
\be
\Pi_S(x-y) = <\frac{g^2}{32\pi^2} F F(x)\frac{g^2}{32\pi^2} F F(y)>,
\ee
containing the action rather than the topological charge, and therefore
counting instantons and anti-instantons  equally. The total topological
charge in a subvolume $V$ is given by
\be
N(V) = \int_V d^4 x \frac{g^2}{32\pi^2} F F(x),
\ee
which satisfies that $\langle N(V) \rangle = n V$.
For pointlike pseudoparticles we have
\be
\frac{g^2}{32\pi^2} F F(x) = \sum_{I=1}^N \delta^4(x - X_I).
\ee
For small $|x-y|$ only terms involving the same instanton contribute
and we find that\footnote{Note that in \cite{glue}
only the correlation function at finite distances $x$ is considered, so the
local term we discuss now is not seen. However, we need it because we
consider the integrated correlator.}

\be
\Pi_S(x-y) = n \delta^4(x-y)\qquad{\rm for }\qquad |x-y|\rightarrow 0.
\label{short}
\ee
In reality the instantons are not pointlike, and show a repulsive interaction.
This leads to the low-energy theorem
\cite{Novikov-etal,Diakonov_Petrov,Ilgenfritz}
\be
\langle [N(V)]^2 \rangle-\langle N(V)\rangle^2 = \frac 4b \langle N(V) \rangle,
\label{low}
\ee
for $V \rightarrow \infty$. The
compressibility of the instanton liquid is thus equal to $4/b$.
As can be seen from (\ref{short}), for small volumes the factor
$4/b$ is absent.  How can we understand this in terms of the physical states
contributing to the correlation function? The lowest lying intermediate state
is  a 'scalar glueball',
which we will generically call $\sigma$ (not to be mixed with the $\sigma$ of
sigma-model, at 600 MeV).
Using a somewhat different convention than
in the pseudoscalar channel the relevant matrix
element is
\be
\langle 0 |\frac{g^2}{32\pi^2} F F|\sigma\rangle = \lambda_\sigma.
\ee
Including the contact term and the contribution of the
continuum with coupling constant $\lambda_c$, this results in the correlator
\be
\Pi_S(x-y) = n \delta^4(x-y) + \lambda_\sigma^2 D(m_\sigma,|x-y|)
+ \lambda^2_c \int_{s_0}^\infty ds D(\sqrt s,|x-y|) s^2.
\label{scalar}
\ee

This result contains both perturbative and nonperturbative
contributions. However, the correlation function that is obtained from the
instanton liquid calculation only receives contributions from the
latter sector. Therefore, we should compare out results to
$\Pi_S^{\rm NP}\equiv \Pi_S - \Pi_S^{\rm pert.}$.
The spectral representation of
the perturbative contribution is given by
\be
\Pi_S^{\rm pert.} = \lambda^2_c \int_{0}^\infty ds D(\sqrt s,|x-y|) s^2.
\label{pert}
\ee

In the same way as for the topological correlator we define
the correlator, $K_S(l_4)$,
as the {\it variance} of the total field
strength inside the box $H(l_4)$. Numerically, this correlator can be
evaluated again by simply counting the number of instantons inside
the box. It is related to $\Pi_S^{\rm NP}$ by
\be
K_S(l_4) = L^3\int_{L^3} d^3 x \int_{-l_4}^{l_4} dt(l_4 - |t|)
\Pi_S^{\rm NP}((\vec x^2 + t^2)^{\frac 12}).
\ee
Using the the correlators (\ref{scalar}, \ref{pert}) we find
\be
K_S(l_4) = n L^3 l_4 + \lambda_\sigma^2L^3(l_4 - \frac{1-
\exp(-m_\sigma l_4)}{m_\sigma}) -\lambda^2_c L^3 \int_0^{s_0}(l_4 -
\frac{1-\exp(-l_4\sqrt s)}{\sqrt s }) s^2 ds.\nonumber\\
\ee
For large volumes the low-energy theorem (\ref{low}) should be obeyed. This
leads to the relation
\be
n + \lambda_\sigma^2 - \lambda_c^2 = \frac 4b n,
\ee
which shows that in the scalar case
(in contrast to pseudoscalar one treated above)
 it is essential to include the contribution
of the continuum. (Recall also that
in the pseudo-scalar case  the sign of contribution from the $\eta'$ resonance
is negative due to the anti-hermiticity of $F\tilde F$.)

In the lower figure of Figs. 2 and 4 and Fig. 5,
we show results for the scalar correlators. The full
lines in all figures correspond to a random positioning of the instantons.
For infinite volume we would get a straight line,  but finite size corrections
turn it into a parabola. The numerical results are represented by dots, and
the dashed line is a fit using the function
\be
f(l_4) = \frac{N}{2L} \left ( \frac 4b l_4(1 - \frac{l_4}{2L}) -
\lambda^2 \frac {1-\exp(-m_\sigma l_4)}{m_\sigma}
+\frac 4{\sqrt s_0} (1+\lambda^2 - \frac 4b)F(l_4\sqrt s_0) \right  ),
\ee
where the 'missing continuum function' $F$ follows from the integral
in (5.10)
\be
F(x) = -\frac 13 + \frac 2{x^3} - (\frac 1x +\frac 2{x^2} +\frac 2{x^3})
\exp(-x)).
\ee
This fitting function both reproduces the short distance behavior (\ref{short})
and the low-energy theorem (\ref{low}).

Results for $N_c =2$ and $N_f = 1$ or $N_f = 2$ are shown in Fig. 4.
Two fits with comparable $\chi^2$ but very different parameters are found.
First, we a fit with $\lambda = 0$ and $\sqrt s_0
= 1110$ MeV for $N_f =1$ and 715 MeV for $N_f = 2$,
respectively. In this case $m_\sigma$ is indeterminate. A second minimum
is found for $\lambda = 1.22$ and $m_\sigma = \sqrt s_0 = 2.4$ GeV for
$N_f = 1$ and $\lambda = 1.32$ and $m_\sigma = \sqrt s_0 = 1.9$ GeV.
In both cases, the
bulk of the $l_4$ behavior is determined by the low-energy theorem, and
the resonance and tcontinuum only contribute significantly for small volumes.

In Fig. 5 we study the current quark mass dependence for $N_c = 3$ and two
flavors of equal mass. Again we find two different fits.
First, for current quark masses equal to 0, 20, 40 and 80 MeV
we find a fit with $\lambda = 0.0$ and
$\sqrt s_0$ equal
to 1160, 1295, 1245 and 1525 MeV, respectively, and, second, we find
$\lambda = $ 1.56, 1.60, 1.44 and 1.72 and
$m_\sigma= \sqrt{s_0} = $ 2.6, 2.9, 2.6 and 3.7 GeV, in the same order. In both
cases, the
bulk of the correlation function is determined by the
low-energy theorem (\ref{low}).

Finally, in the lower figure of Fig. 3, we show $K_S$ for realistic values
of the quark masses (see caption). As a best fit with $\lambda = 0$
we find that $\sqrt s_0$ = 1320 MeV.
For the fit with $m_\sigma= \sqrt{s_0}$ we find a mass of 2.97 GeV and
a coupling constant of $\lambda = 1.53$.
These results are consistent with the factor $4/b$ in the low-energy
theorem.

We want to emphasize that the value $\lambda = 0$ and the  equality
$m_\sigma= \sqrt{s_0}$ are a result of the fit. The fits quoted here are
not the only fits; there are other minima of $\chi^2$ which we dismissed as
unphysical. For example, in the last case is also fitted by $\sqrt{s_0} =
\frac 43 m_\sigma = 200$ MeV and $\lambda=4.5$. For this ratio of the masses
the asymptotics of the fitting function becomes independent of $\lambda$.
Therefore the conclusion that the bulk of the results are described
by the low-energy theorem remains unaffected.

\vskip 1.5cm
\section{Conclusions}
\vskip 0.5cm

We have presented a detailed study of
the volume dependence of the fluctuations of the
topological charge in an Interaction Instanton Approximation to the QCD vacuum.
As was
expected, we found that these fluctuations are screened for a sufficiently
large box and sufficiently small quark masses due to the
fermion-induced interaction between
instantons. The phenomenon disappeared gradually as the quark mass
increases from zero. An upper limit for the Samuel's critical mass value is
set to about 20 MeV.
  Furthermore, the results for
a {\it correlated} instanton liquid have shown that correlations are
essential for reproducing the $m_\eta'$ and its dependence on the number of
colors and flavors in the theory.

As a by-product, we have obtained a formula similar to
the Witten-Veneziano
formula, but  the topological susceptibility that enters into it
is not the large volume limit but rather the small volume limit in
the presence of dynamical fermions.

  We also present results for the fluctuations of total number of instantons
in the finite box. The results clearly indicate that the instanton liquid is
not an ideal gas, but a liquid with an important repulsive interaction. The
results agree very well with the NSVZ low-energy theorem, and provide
constraints for  the parameters of the physical particles
in the scalar channel.

Finally, let us emphasize that the method  to evaluate the $\eta'$ parameters
proposed in this work can be applied to lattice QCD calculations with
dynamical fermions as well. Straightforward evaluation
of the quark propagators can be substituted by measurements of the topological
charge in a finite box, which may be much easier to do.

\vskip 1.5cm
\noindent
{\bf{\large Acknowledgments}}
\vskip 0.5 cm
This work was supported in part by the US department of energy under Grant
DE-FG02-88-ER40388. We acknowledge the NERSC at Lawrence Livermore where
most of the computations presented in this paper were performed.
Th. Sch\"afer and I. Zahed are acknowledged for useful discussions.

\vskip 1.5cm
\noindent
{\bf{\large Appendix A: Topological correlators in effective field theory}}
\vskip 0.5cm
\renewcommand{\theequation}{A.\arabic{equation}}
\setcounter{equation}{0}

The topological charge density is defined by
\be
Q(x) = \frac{g^2}{32\pi^2} F \tilde F(x),
\label{QX}
\ee
where $F$ is the field strength tensor. The topological charge $Q(V)$ in
4-volume $V$ is
obtained by integration of $Q(x)$ over $V$. The fluctuations of the
topological charge in a given box are determined by
the topological correlation function defined by
\be
\Pi_P(x-y) = \langle\frac{g^2}{32\pi^2} F \tilde F(x)\frac{g^2}{32\pi^2}
F \tilde F(x)\rangle.
\label{KP}
\ee
In this appendix we will derive this correlator from
the effective low-energy chiral Lagrangian \cite{Div} and its interaction
with the topological charge $Q$. This derivation is based on work in
references \cite{THOOFT-1986,DOWRICK-MCDOUGALL,KIKUCHI-WUDKA-1992}.
The main additional ingredient is that we include the $\eta-\eta'$ mixing
in our derivation.

The topological charge density couples to the pseudoscalar singlet
meson channel, and, for a diagonal mass matrix with $m_u = m_d$, only the
$\eta-\eta'$ mixing part
of the effective pseudoscalar Lagrangian is of
relevance\footnote{In fact, there is another state
with the quantum numbers of the $\eta'$,
$\eta(1430)$ and formerly called $\iota$, which gives a comparable
contribution to this correlation function (see discussion
in \cite{Shuryak_cor}). We do not include it here,
because it is hardly possible to
extract any information about this state from the data sample discussed
in this paper.}.
This part, which
involves only the diagonal $SU(3)-$flavor fields $\phi_0$ and $\phi_8$,
is given by \cite{Veneziano-1979,Div,dp-1986,NVZ,NVZ-chir}
\be
{\cal L}_{\rm eff} = -i \int d^4 x \frac{\sqrt {2N_f}}{f} \phi_0 Q +
{\cal L}(\phi_0, \phi_8),
\label{eff1}
\ee
with
\be
{\cal L}(\phi_0, \phi_8) = \frac 12 \left (
\begin{array}{c} \phi_0 \\ \phi_8 \end{array}\right )
\left ( \nabla^2 + {\cal M}^2 \right )
\left (\begin{array}{c} \phi_0 \\ \phi_8 \end{array} \right ),
\label{eff2}
\ee
and the square of the mass matrix \cite{Veneziano-1979}
\be
{\cal M}^2 = \left ( \begin{array}{cc} \frac 23 m_K^2 + \frac 13 m_\pi^2 &
\frac{2\sqrt 2}3(m_\pi^2 - m_K^2) \\ \frac{2\sqrt 2}3(m_\pi^2 - m_K^2) &
\frac 43 m_K^2 - \frac 13 m_\pi^2
\end{array} \right ).
\label{eff3}
\ee
The normalization of the fields and the coupling constant $f$ will
be discussed below.

In the dilute gas approximation the instanton partition function is given by
\cite{CDG,THOOFT-1986}
\be
Z = \sum_{N_+\, N_-} \frac{\kappa^{N_+ +N_-}}{N_+! N_-!}
\prod_{i=1}^N d^4 z_i
\exp(-\int d^4x {\cal L}_{\rm eff}).
\label{eff4}
\ee
The constant $\kappa$ was calculated to one loop order in \cite{THOOFT-1976}.
Treating the instantons as pointlike objects, the topological charge density
is given by $Q(x) = \sum Q_i \delta(x - z_i)$. This allows us to perform the
sum over $N_+$ and $N_-$ in the partition function. Our final effective
Lagrangian is obtained by expanding
the resulting cosine-function to second order (weak field approximation):
\be
{\cal L}_{\rm eff} = -2V \kappa + \kappa \frac{2N_f}{f^2} \phi_0^2 +
{\cal L}(\phi_0, \phi_8).
\label{eff5}
\ee
The constant $\kappa$ can be traded for the average pseudoparticle density
by differentiating both (\ref{eff4}) and (\ref{eff5}) with respect to
$\kappa$.
We find
\be
2 \kappa = \langle\frac NV\rangle\equiv n.
\label{eff6}
\ee

The topological correlator (\ref{KP}) can be obtained by integrating out
the meson fields.
However, it is a somewhat simpler to write the correlator
in terms of functional derivatives
with respect to $\phi_0$ of the first term in the Lagrangian (\ref{eff1})
and performing the differentiation after the sum over the instantons, i.e. on
the second term in (\ref{eff5}). The correlator can be expressed as
\be
\langle Q(x) Q(y) \rangle = n \delta^4(x-y) - n^2
\frac{2N_f}{f^2} \langle \phi_0(x) \phi_0(y) \rangle.
\label{eff7}
\ee
The latter expectation value is most conveniently obtained by
diagonalizing the
mass matrix ${\cal M}$ (including the topological contribution)
with eigenvalues identified as $m_{\eta'}$ and $m_\eta$ and eigenvectors
expressed in the $\eta'-\eta$ mixing angle according to
\be
|\eta'> &=& \cos\phi |\phi_0> + \sin\phi |\phi_8> ,\\
|\eta> &=& \cos\phi |\phi_8> - \sin\phi |\phi_0> .
\label{eff8}
\ee
For the singlet correlation function we obtain
\be
\langle \phi_0(x) \phi_0(y) \rangle = \cos^2(\phi) D(m_{\eta'} |x-y|) +
\sin^2(\phi) D(m_{\eta} |x-y|)
\label{eff9}
\ee
resulting in the topological correlator
\be
\langle Q(x) Q(y) \rangle = \frac {f^2}{2N_f}
m^2_{\rm top} \left [ \delta^4(x-y) - m^2_{\rm top}(\cos^2(\phi)
D(m_{\eta'}, |x-y|) +\sin^2(\phi) D(m_{\eta}, |x-y|) ) \right ],
\label{eff10}\nonumber\\
\ee
where the scalar propagator is given by $D(mx) = m K_1(mx)/(4\pi^2 x)$.
For convenience we have introduced the topological mass
\be
m_{\rm top}^2 \equiv \frac {2N_f n}{f^2}=m_{\eta'}^2 + m_\eta^2 - 2m_K^2,
\label{eff11}
\ee
where the latter identity follows from the invariance of the trace
of the mass matrix under diagonalization. This correlation function can
also be derived from the pseudoscalar singlet spectral function with the
matrix element $\langle 0 | Q(x)| \eta' \rangle = m_{\eta'}^2 f_{\eta'}/
\sqrt{2N_f}$. This defines the constant and the normalization of the
fields in our effective Lagrangian.

It is interesting to calculate the topological susceptibility
\be
\chi \equiv \lim_{V\rightarrow\infty}
\frac {\langle (\int d^4 x Q(x) )^2 \rangle}V =
\frac{f^2}{2N_f} m_{\rm top}^2\left
 ( 1- \frac{(\frac 43 m_K^2 - \frac 13 m_\pi^2)
m_{\rm top}^2}{(\frac 43 m_K^2 - \frac 13 m_\pi^2)
m_{\rm top}^2 +2m_K^2 m_\pi^2 - m_\pi^4}\right).
\label{eff12}
\ee
We observe the well-known fact that the topological charge is completely
screened if $one$ massless quark is present. Note that this formula has been
derived for $m_u=m_d$. Indeed, for zero light quark masses
we have $m_{\pi} = 0$ and
$\chi = 0$. If $m_s = 0$ we have that $m_\pi^2 = 2 m_K^2$ and also $\chi = 0$.

As has been particularly emphasized by Dowrick and McDoughal
\cite{DOWRICK-MCDOUGALL}, we can look at the effective
Lagrangian in a different way. Namely we integrate over the pseudoscalar
singlet
and are left with a residual instanton interaction. In the present case,
with the $\eta'-\eta$ taken into account, we have to integrate over
both $\phi_0$ and $\phi_8$.
This results in the effective lagrangian
\be
{\cal L} = \frac 12 \frac{2N_f}{f^2}\sum_{ij} Q_i Q_j \left(\cos^2\phi
D(m_\pi|z_i-z_j|)+\sin^2\phi D(m_K|z_i-z_j|)\right).
\label{eff13}
\ee
The range of the interaction between the pseudoparticles is of
order of $1/m_\pi$. Ignoring the
the $\eta'-\eta$ mixing would have led to the improper conclusion \cite{Ismail}
that interaction range of $1/m_K$.
Although it is believed that a Yukawa interaction
is not strong enough \cite{Shrock} to induce a plasma phase, the
relatively long range interaction justifies the study of the
scenario recently proposed by Samuel \cite{SAMUEL-1992}.

\newpage
\setlength{\baselineskip}{15pt}

\vfill
\newpage
\noindent
{\bf Figure Captions}
\vskip 0.5 cm

\noindent
Fig. 1. The integrated topological charge in the Debije cloud of an
anti-instanton in a slice
of length $l_4$ in the 4 direction of total length $2L$. The results
are for $N_c = 3$ and two flavors with masses (in units of $fm^{-1}$
as given in the label of the figure.

\vskip 0.5 cm
\noindent
Fig. 2. The pseudoscalar correlator $K_{P}(l_4)$ (upper figure) and the
scalar gluonic correlator $K_S(l_4)$ (lower figure) as a function of the
length of the slice, $l_4$, in the 4 direction of total length $2L$. The
full line is for randomly positioned instantons and the dashed curves
represent results of a fit. All result in this figure are for two colors and
one or two massless flavors.

\vskip 0.5 cm
\noindent
Fig. 3. The mass dependence of the pseudoscalar topological correlator for
three colors and two flavors of equal mass (in units of $fm^{-1}$
given in the label of the figures.
For further explanation we refer the the caption of Fig. 2.

\vskip 0.5 cm
\noindent
Fig. 4. The pseudoscalar correlator $K_{P}(l_4)$ (upper figure) and the
scalar gluonic correlator $K_S(l_4)$ (lower figure) as a function of the
length $l_4$ of the subvolume
for $N_c =3$, $N_f = 3$ and $m_u = m_d = 10 MeV$ and $m_s = 150 MeV$.
Further explanation can be found in the caption of Fig. 2.

\vskip 0.5 cm
\noindent
Fig. 5. The mass dependence of the scalar topological correlator, $K_S(l_4)$,
for
three colors and two flavors of equal mass (in units of $fm^{-1}$)
given in the label of the figures.
For further explanation we refer the the caption of Fig. 2.


\begin{thebibliography}{10}
\bibitem{Novikov-etal} V. Novikov, M. Shifman, A. Vainshtein and V. Zakharov,
Nucl. Phys. {\bf B165} (1980) 67.

\bibitem{Polyakov_etal}A.A. Belavin, A.M. Polyakov,  A.A. Schwartz
and Yu.S. Tyupkin, Phys.Lett. {\bf 59B} (1975) 85.

\bibitem{THOOFT-1976} G. 't Hooft, Phys. Rev. Lett. {\bf 37} (1976) 8; Phys.
Rev. {\bf D14} (1976) 3432.

\bibitem{I}E.V.  Shuryak and J.J.M. Verbaarschot, Nucl. Phys. {\bf B 410}
(1993) 37.

\bibitem{II}E.V.  Shuryak and J.J.M. Verbaarschot, Nucl. Phys. {\bf B 410}
(1993) 55.

\bibitem{III}T. Sch\"afer,
E.V.  Shuryak and J.J.M. Verbaarschot, Nucl. Phys. {\bf B 412} (1994) 143.

\bibitem{Shuryak_cor}
E.~V.~Shuryak, \newblock Rev. Mod. Phys. {\bf 65}, 1 (1993)

\bibitem{Negele_etal}
M.C.~Chu, J.M.Grandy, S.Huang, and J.W.Negele,
\newblock  Phys. Rev. Lett. {\bf 70}, 255 (1993)


\bibitem{Schafer-Shuryak}
T.~Sch\"afer, E.~V.~Shuryak, Hadronic Wavefunctions in the Instanton
model, preprint, SUNY-NTG94-5, hep-ph/9401289.

\bibitem{cooling}
B.~Berg, Phys.~Lett.~{\bf B114}, 475 (1981);
M.~Teper, Nucl.~Phys.~{\bf B20} (Proc.~Suppl.), 159 (1991).

\bibitem{Chu_etal_2}
M.~C.~Chu, J.~M.~Grandy, S.~Huang , J.~W.~Negele,
preprint, MIT-CTP\#2269, hep-lat/9312071

\bibitem{Shuryak_82}
E.~V.~Shuryak,
\newblock Nucl. Phys. {\bf B203}, 116, 140, 237 (1982)

\bibitem{T} T. Sch\"afer,
E.V.  Shuryak and J.J.M. Verbaarschot, Nucl. Phys. {\bf B 412} (1994) 143.

\bibitem{SV_next}T. Sch\"afer,
E.V.  Shuryak and J.J.M. Verbaarschot, {\it Mesons in the  correlated
instanton vacuum}.

\bibitem{PECCEI-QUINN-1977}R. Peccei and H. Quinn",
Phys. Rev. Lett. {\bf 38} (1977) 1440

\bibitem{Schierholtz} G. Schierholtz, {Towards a dynamical solution of the
strong CP problem}, DESY preprint DESY 94-031 (1994).

\bibitem{SAMUEL-1992}S. Samuel,  Mod. Phys. Lett. {\bf A7} (1992) 2007.

\bibitem{Weinberg} S.Weinberg, Phys. Rev. {\bf D11} (1975) 3583.

\bibitem{THOOFT-1986}G. 't Hooft, Phys. Rep. {\bf 142} (1986)357.

\bibitem{DOWRICK-MCDOUGALL}N.J. Dowrick and N.A. Mcdougall,
Phys. Lett. {\bf B285} (1992) 269; Nucl. Phys. {\bf B399}(1993) 426.

\bibitem{KIKUCHI-WUDKA-1992} H. Kikuchi and J. Wudka,
Phys. Lett. {\bf B284} (1992) 111.

\bibitem{Ismail} I. Zahed, {\it QCD instantons in vacuum and matter},
Stony Brook preprint SUNY-NTG-94-22.


\bibitem{Witten-1979} E. Witten, Nucl. Phys. {\bf 156} (1979) 269.

\bibitem{Veneziano-1979} G. Veneziano, Nucl. Phys. {B159} (1979) 213.

\bibitem{Jan-Smit}J. Smit and J. Vink, Nucl. Phys. {\bf B284} (1987) 234.

\bibitem{SMILGA} A. Smilga, Phys. Rev.  {\bf D 46} (1992) 5598.

\bibitem{Ukawa} Ukawa, Phys. Rev. Lett. {\bf 72} (1994) 3448.

\bibitem{glue}
T.~Sch\"afer, E.~V.~Shuryak, $Glueballs and Instantons$,  SUNY-NTG preprint.

\bibitem{Shuryak_1988}
E.~V.~Shuryak,
\newblock Nucl. Phys. {\bf B302}, 559,574,599 (1988)

\bibitem{Shuryak_1989}
E.~Shuryak,
\newblock Nucl. Phys.{\bf B319}, 511,521 (1989); {\bf B328}, 85,102 (1989)

\bibitem{SHURYAK-VERBAARSCHOT-1991} E.V. Shuryak and J.J.M. Verbaarschot,
Nucl. Phys. {\bf B364} (1991) 255.

\bibitem{Diakonov_Petrov}
D.~I.~Diakonov and V.~Yu.~Petrov,
\newblock Nucl. Phys. {\bf B245}, 259 (1984); D.~I.~Diakonov, V.~Yu.~Petrov,
Nucl.~Phys.~{\bf B272}, 457 (1986).

\bibitem{NVZ}
M.~A.~Nowak, J.~J.~M.~Verbaarschot and I.~Zahed,
\newblock Nucl. Phys. {\bf B324}, 1 (1989)

\bibitem{VERBAARSCHOT-1991} J.J.M. Verbaarschot, Nucl. Phys. {\bf B362} (1991)
33.

\bibitem{SHURYAK-VERBAARSCHOT-1992A} E.V. Shuryak and J.J.M. Verbaarschot,
Phys. Rev. Lett. {\bf 68} (1992) 2576.

\bibitem{Yung}A.V. Yung, Nucl. Phys. {\bf B297} (1988) 47.

\bibitem{Diakonov-Petrov-1993} D.~I.~Diakonov, V.~Yu.~Petrov,
Nucl.~Phys. (1993).

\bibitem{Langfeld-Reinhardt} K. Langfeld and H. Reinhardt, {Scale anomaly
induced instanton interaction} (1994).

\bibitem{Div}P. Di Vecchia and G. Veneziano, Nucl. Phys. {\bf B171} (1980) 253.

\bibitem{dp-1986}D.I. Diakonov and V.Yu. Petrov,
Leningrad Preprint LNPI-1153, 1986.


\bibitem{NVZ-chir}
M.A. Nowak, J.J.M. Verbaarschot and I. Zahed,
Phys. Lett. {\bf B228} (1989) 251; R. Alkofer, M.A. Nowak,
J.J.M. Verbaarschot and I. Zahed, Phys. Lett. {\bf B233} (1989) 205.

\bibitem{CDG}
C.~G.~Callan, R.~Dashen and D.~J.~Gross,
Phys.~Rev.~{\bf D17}, 2717 (1978),

\bibitem{Shrock} R. Shrock, $Lattice Yukawa models$ in $Quantum Fields on
the Computer$, ed. M. Creutz, World Scientific 1992.

\bibitem{Ilgenfritz} E.M. Ilgenfritz, Habilitation Leipzig, 1988.

\end{thebibliography}
\end{document}